# A Monte Carlo Simulation study of Most Likely Position (MLP) and Position Vector (PV) methods in TOFPET

Nagendra Nath Mondal [*,1]

[1] Saha Institute of Nuclear Physics, 1/AF Bidhannagar, Kolkata 700064, India.

**Abstract**

The results of Monte Carlo Simulation (MCS) studies of Most likely position (MLP) and position vector (PV) methods in TOFPET system are presented. MCS based on GEANT3.21 is carried out where the geometry of a real TOFPET system is considered. Results not only manifest resolving powers ($R_P$) of PV and MLP methods ~114% and ~36% but also exhibit shifting of reconstructed images from the original positions ~3% and ~63% respectively. Position conversion factors play a crucial role to reinstate the image position in the PV method and stipulate excellent images. A PV is a position reconstruction method of positron-electron annihilation points developed afresh without iteration and that makes its beauty by saving huge computational time and radiation dose of the patient.

**1 Introduction** An image reconstruction in Time-of-Flight Positron Emission Tomography (TOFPET) is a state-of-art technique. So far interest of improving image quality with less blurring effects, backgrounds and artifacts a short time interval (typically 4-8 ns for a human PET) between two back-to-back (btb) collinear photons originated from the positron-electron ($e^+e^-$) annihilation point is usually set and data are collected in coincidence mode. The limitation of timing resolution is about 4 ns in the current PET technology constrains the position to a 60 cm region indicates that the measurement and reconstruction techniques are not good enough to reproduce the original position. Consequently Vandenberghe and Karp [1] selected coincidence events with more precision between btb collinear photons in a line of response (LOR) of TOFPET. Fourier back projection, Maximum-likelihood [2] and LU decomposition [3] are a few excellent image reconstruction techniques. Most likely position (MLP) is a conventional image reconstruction technique in TOFPET [1], where $e^+e^-$ annihilation points are reconstructed inside the LOR by compelling 511 keV with 4 ns timing window. All of these techniques are based on iteration. 'Position Vector (PV)' is a recently developed image reconstruction method without iteration in TOFPET [4], where annihilation events are selected by energy-time (E-T) correlation technique. A Monte Carlo Simulation (MCS) based on GEANT3.21 [5] is executed in order to compare basic differences, resolving powers, advantages and inconveniences between MLP and PV methods. In the following sections MCS of TOFPET, image reconstruction of MLP and PV methods, results, discussions and finally conclusions are presented.

[*] Corresponding author: e-mail nagendra.mondal@saha.ac.in, Phone: +91 33 2337 5345, Fax: +91 33 2337

**2 Monte Carlo Simulation of TOFPET** An extensive MCS based on GEANT3.21 is adopted for the development of an exquisite TOFPET. The TOFPET system consists of 96 detectors in two rings. Ingredients of each detector are a cylindrical $Lu_2SiO_5$:Ce (in short LSO, size $3\varphi \times 3$ cm$^3$ ) crystal and a cover of PMT (Iron brass of size $3\varphi \times 12$ cm$^3$). Detectors are arranged on the x-y plane in perpendicular to the z-axis. A similar layout of TOFPET system (single ring) can be found elsewhere [6], where 48 detectors were consisted of cylindrical $BaF_2$ crystals ($1.5\varphi \times 2$ cm$^3$). In spite of lower time resolution of LSO than $BaF_2$ it is considered because of the highest γ-ray detection efficiency and atomic density. The size of LSO is optimized and it is covered by a thin Al (0.01 cm) and a thick Pb (0.5 cm) shields in order to reflect scintillation light and stop Compton scattered γ-rays. The diameter of the ring is 80 cm. In Figure 1 a simulated TOFPET system is depicted.

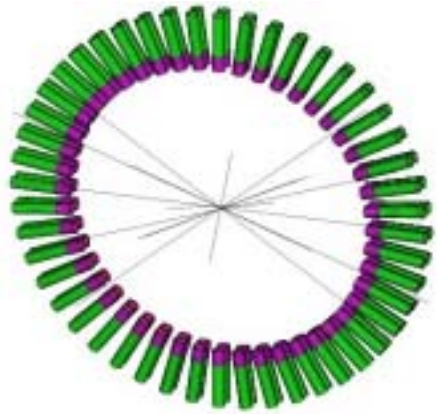

**Figure 1** An array of 96 detectors in two rings of a standard TOFPET system. Annihilation points and btb collinear γ–rays are shown clearly.

Energy and time resolutions of LSO are incorporated in the MCS. Both energy and time information of each detected photons are recorded in different channels (assuming 96 ADC and 96 TDC channels) in the data acquisition system.

**3 Position Reconstruction of $e^+e^-$ annihilation points** In this study more attention is given to the MLP and the PV methods for comparison between the reconstructed images. $e^+e^-$ annihilation events (each event consists of two btb collinear 511 keV γ–rays) are generated and energy and time of each detected γ–rays are recorded. Those data are analyzed in the following two methods.

**3.1 Most Likely Position (MLP) method** It is a real time method without accounting the angular dependence of the detectors. The accuracy of the time information is assumed to be good enough to assign all activity directly to one point in the matrix [1]. A LOR is considered whose end points are $(x_1,y_1,z_1)$ and $(x_{25},y_{25},z_{25})$ and a TOF difference $\Delta t = t_{25} - t_1$ of a single pair in the same ring. The reconstructed positions in 2D is given by

$$x_{MLP} = \frac{(x_1 + x_{25})}{2} - \frac{C\Delta t}{2}\frac{(x_{25} - x_1)}{d} \qquad (1)$$

and

$$y_{MLP} = \frac{(y_1 + y_{25})}{2} - \frac{C\Delta t}{2}\frac{(y_{25} - y_1)}{d} \qquad (2)$$

Where $C$ is the speed of light, and $d = \sqrt{(x_{25}-x_1)^2 + (y_{25}-y_1)^2 + (z_{25}-z_1)^2}$ is the length of the LOR. The first term of each equation (1 and 2) is zero (see Figure 1), so that image is produced only by the second term. The reconstructed positions ($x_{MLP}, y_{MLP}$) of the other pairs and second ring are obtained by similar techniques with respective coordinates of the detector.

**3.2 Position Vector (PV) method** In this case position of the detector is determined by the TOF and speed of the incident photons on the corresponding detector. The position vectors of $X$- and $Y$- components of the face-to-face detectors are determined from the respective TOF and $C$, and the difference of those positions determines the annihilation point of $e^+e^-$. Details of the method are available elsewhere [4]. The generalized forms of equations of position reconstruction in 2D are given by

$$X = \sum_{i=1}^{n/2}\{([T_i|\cos(i-1)\theta| - T_{i+n/2}|\cos(i+n/2-1)\theta|] \times C/2) - c\}/2m \qquad (3)$$

and

$$Y = \sum_{i=1}^{n/2}\{([T_i|\sin(i-1)\theta| - T_{i+n/2}|\sin(i+n/2-1)\theta|] \times C/2) - c\}/2m \qquad (4)$$

where $T_i$ and $T_{i+n/2}$: TOF of btb photons of face-to-face detectors, $\theta = 2\pi/n$: angle between the neighboring detectors, $n$: even number of detectors per ring. $c, m$: respectively are the position conversion factors (intercept and slope respectively) which are introduced in order to reinstate the image position [4]. Equations 1–4 are the case of a single ring and for multiple rings the second term will be changed accordingly to make the cross LOR which is taken into account in the image reconstruction process.

**4 Results and Discussions** We have generated equal number of $e^+e^-$ annihilation events (diameter of the source area is 1 cm) at the two different positions (2,0,0) and (-2,0,0) of a TOFPET system (see Figure 1). Image is reconstructed in every LORs (including cross detectors alignment between two rings) of the system by MLP method (sec. 3.1), images of every LORs are added and depicted in Figure 2.

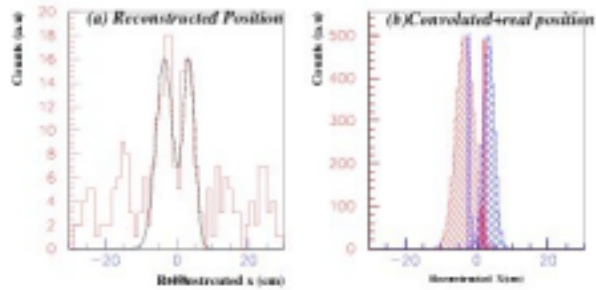

**Figure 2** (a) Reconstructed image of two positions along x-axis. (b) Their convoluted images are overlapped with their original distributions. A large dark zone is observed.

During the reconstruction process the 4 ns time and 450 –550 keV energy windows are given as a constrained. From the spectra it can be observed that a lot of backgrounds (Figure 2a) are appeared in both sides beyond 10 cm of the main peaks. Although no natural backgrounds are generated in the MCS, but a huge number of those events are seen in the reconstructed images. In the TOF data time of multiple Compton scattering of incident photons inside the scintillator are cumulated.

Therefore difference of TOF produced in the MLP method contains several peaks of scattered photons. It is difficult to suppress or isolate those images from the real one except very tight energy and time window settings. Giving the 1 ns timing window it is observed that backgrounds are reduced but not removed completely. In Figure 2b a convoluted spectrum (produced from the fitting parameters of Figure 2a) is presented with their original position distributions. Spectra show that the shifting of reconstructed positions from the original positions is ~63% in average and resolving power ($R_P$: measures the minimum resolvable distance between the two distinguishable objects, $R_p = (\sigma_{avor} / \Delta\sigma_{rec}) \times 100\%$, where σ is the uncertainty of the Gaussian distribution) is ~36%. More than one source position inside the crossing area of the two spectra ("dark zone" around 8 cm) is impossible to resolve by this method.

Similarly, we have reconstructed images of the same data by the PV method where E-T correlation technique is applied with 1 ns timing and 950-1050 keV energy windows. Images are depicted in Figure 3. No backgrounds can be seen beyond 4 cm from the reconstructed images (3a).

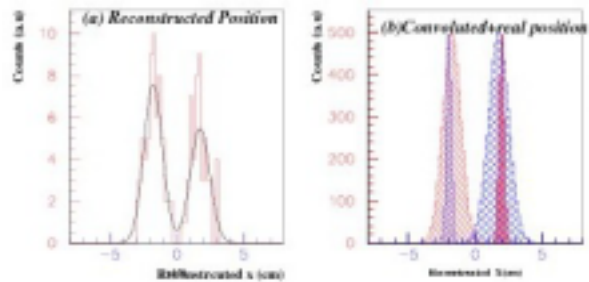

**Figure 3** (a) Reconstructed images of two positions; and (b) their convoluted images with original distributions are shown. Reconstructed positions are shifted slightly and dark zone is negligible.

Convoluted spectrum is produced as before attributes less blurring effects, no significant dark zone and the deviation from the original spectra is only 3%. Consequently, position conversion factors play a vital role to reinstate the reconstructed position. The $R_P$ is obtained to be ~114% and it implies that two adjacent small size tumors of diameter 1 cm can easily be resolved by this technique. The dark zone of PV is eight times smaller than that of MLP method. The higher $R_P$ shows better image quality, i.e., less blurring effects in the image. $R_P$ of PV is about three times larger than the MLP method. PV method improves the image quality in the true sense that real events are selected by E-T correlation technique and narrow timing window restricts multiple scattering times inside the scintillator.

It is hard to distinguish the original source positions from the image produced by MLP method, but PV method is succeeded to do that. MLP is an iterative technique requires much time (1iteration of 1 million events takes ~50 sec), huge data (~GB) as well as a few hundred of mCi dose in the patient body, and large memory of the computer for data storing and analysis. PV is a non-iterative and online image reconstruction technique can avoid such inconveniences of MLP method.

**5 Conclusions:** A comparative study between MLP and PV image reconstruction methods is done by MCS based on GEANT3.21 in a TOFPET system. PV method shows better performance than MLP method i.e., minimum blurring affect, better resolving power and least backgrounds in the reconstructed image. Image of the PV method is better matched with the original distributions with lower detection efficiency than the MLP method, but MLP suffers from the positioning problem and huge backgrounds. Only a few millions events are good enough to reproduce a high quality image without any iteration in PV method. Advantages of this method are the reduction of dose and savings of computational time.

**Acknowledgements:** Author is pleased to acknowledge the Visiting fellowship supports from the SINP and the VECC, and Computer facility of the INO project.